\documentclass[aps,pre,twocolumn]{revtex4}
\usepackage{amssymb}
\usepackage{amsmath}
\input{epsf}

 \def\Xint#1{\mathchoice
{\XXint\displaystyle\textstyle{#1}}%
{\XXint\textstyle\scriptstyle{#1}}%
{\XXint\scriptstyle\scriptscriptstyle{#1}}%
{\XXint\scriptscriptstyle\scriptscriptstyle{#1}}%
\!\!\int}
\def\XXint#1#2#3{{\setbox0=\hbox{$#1{#2#3}{\int}$ }
\vcenter{\hbox{$#2#3$ }}\kern-.5\wd0}}

\def\dashint{\Xint-}

\def\R{\mathbb{R}}

\def\bq{\begin{equation}}
\def\eq{\end{equation}}
\def\bqy{\begin{eqnarray}}
\def\eqy{\end{eqnarray}}

\def\De{\Delta}
\def\wi{M}

\begin{document}

\title{Undamped electrostatic plasma waves}

\author{F. Valentini$^1$, D. Perrone $^1$, F. Califano$^2$, F. Pegoraro$^2$,
P. Veltri$^1$, P. J. Morrison$^3$ and T. M. O'Neil$^4$}
\affiliation{$^1$Dipartimento di Fisica and CNISM, Universit\`a della Calabria,
87036 Rende (CS), Italy \\ $^2$Dipartimento di Fisica and CNISM, Universit\`a di
Pisa, 56127 Pisa, Italy\\ $^3$Institute for Fusion Studies and Department of Physics, University of Texas at
Austin, Austin, TX 78712-1060\\
$^4$Department of Physics, University of California at San Diego, La Jolla,
California, 92093}


\begin{abstract}
Electrostatic waves in a collision-free unmagnetized plasma of electrons with fixed ions are investigated for electron equilibrium velocity distribution functions that deviate slightly from  Maxwellian.  Of interest are undamped waves that are the  small amplitude limit of nonlinear excitations, such as electron acoustic waves (EAWs).   A deviation consisting of a small plateau, a region with zero velocity derivative over  a width that is  a very small fraction of the electron thermal speed,  is shown to give rise to new undamped modes, which here are named  {\it corner modes}.  The presence of the  plateau  turns off  Landau damping and allows oscillations with phase speeds within the plateau.   These undamped  waves are obtained in a  wide region of the $(k,\omega_{_R})$ plane ($\omega_{_R}$ being the real part of the wave frequency and $k$ the wavenumber), away from  the well-known `thumb curve' for Langmuir waves and EAWs based on the Maxwellian.  Results of nonlinear Vlasov-Poisson simulations  that corroborate the existence of these modes are described.  It is also shown that deviations caused by fattening the tail of the distribution shift roots off of the thumb curve toward lower $k$-values and chopping the tail shifts them toward  higher   $k$-values. In addition,  a rule of thumb is obtained for assessing how the existence of a plateau shifts roots off of the thumb curve.   Suggestions  are made for  interpreting  experimental observations of electrostatic waves,  such as recent ones in nonneutral plasmas.
\end{abstract}

\date{\today}
\maketitle

\section{Introduction}
\label{intro}

In his 1946 seminal paper \cite{landau46} Landau  demonstrated that electrostatic plasma waves of vanishing
amplitude can be damped, due to their interaction with particles that stream with velocities close
to the wave phase speed,  $v_\phi$.  For unmagnetized uniform plasmas, the wave damping  rate is generally
proportional to the slope of the equilibrium distribution of particle velocities  at $v_\phi$.
Therefore, for monotonically  decreasing equilibrium velocity distribution functions  (such as the usual Maxwellian)
plasma waves are damped exponentially in time.

Almost twenty years later O'Neil \cite{oneil65} analyzed the effects of nonlinearity on the propagation
of plasma waves and found that the process of particle trapping in the wave potential well can inhibit
Landau damping, by flattening the velocity distribution near  the wave phase speed.

In 1991 Holloway and Dorning \cite{holloway91} noted that certain nonlinear electrostatic oscillations can survive
Landau damping even when their phase velocities are comparable to the electron thermal speed, $v_{th}$,
due to the effect of particle trapping. They called these waves electron acoustic waves (EAWs),
since in the range of small wavenumbers their dispersion relation is of the form $\omega\simeq 1.31
kv_{th}$. An EAW is in fact a Bernstein-Greene-Kruskal (BGK) mode \cite{bernstein57} with a  velocity
distribution function that is effectively flat at the wave phase speed, due to trapping.  In 1991, Demeio and
Holloway \cite{demeio91} performed Vlasov-Poisson simulations and provided evidence for the analytical results of
Ref.~\cite{holloway91}.   Also in Ref.~\cite{holloway91}, the authors  claimed that these undamped plasma oscillations
have no  linear counterpart, since in their construction the trapped particle distribution vanishes with wave amplitude and approaches  a Maxwellian, for which the waves are heavily damped.

The latter point was discussed in 1994 by Shadwick and Morrison \cite{shadwick94},  where it was pointed out that  stationary inflection point modes  are the natural linear  limit of the EAWs (BGK modes) of  Ref.~\cite{holloway91}.  A  stationary inflection point  mode  is  an undamped  mode with phase speed $v_\phi$ at  a point where the first two derivatives of a stable homogeneous equilibrium velocity distribution function vanish, and by the Penrose criterion this is necessary for the simultaneous existence of the mode and stability.  The issue here is that the  homogeneous equilibrium distribution function  approached as the wave  amplitude approaches zero is not unique,  and  there is no {\it a priori} reason it should be locally  Maxwellian,   since the structure  near $v_{\phi}$ is determined by the history of formation of the waves.  Given a nonlinear BGK wave, there are many ways the  limit of the vanishing of trapped particles can be taken  and   the result is clearly norm dependent. (See Ref.~\cite{hagstrom} for a discussion of bifurcations in $W^{1,1}$.)

The existence of the EAW branch has been investigated in many electrostatic Particle-In-Cell \cite{valentini06}
and Vlasov-Poisson \cite{afeyan04,johnston09} simulations. Moreover, recent Vlasov-Yukawa
simulations (with Vlasov ions and linear adiabatic electrons)  led to the  prediction of undamped electrostatic waves at
low frequencies (of the order of the proton plasma frequency), similar in nature to the EAWs and dubbed ion-bulk waves
\cite{valentini111,valentini112}.   EAW-type fluctuations have been detected in spacecraft
data from observations in the interplanetary medium \cite{gary85}. Also, the excitation of the EAWs has been
obtained in laboratory experiments with nonneutral plasmas \cite{anderegg091,anderegg092}, in which an
external driving electric field is applied to the plasma column for the time needed to create a population
of trapped particles, with the waves surviving after the drive is turned off. The flat region in the particle velocity distribution generated by trapping at the wave phase speed inhibits Landau damping and allows  the EAWs to survive.   The experimental results discussed in Refs.~\cite{anderegg091,anderegg092} confirm the existence of EAWs on the nonneutral analog of the so-called thumb curve of  Ref.~\cite{holloway91}, as is the case for the numerical results in Refs.~\cite{afeyan04,valentini06,johnston09}; however,  these experiments also suggest  that  wave excitation can be obtained off of  the usual  thumb curve of  Ref.~\cite{holloway91}, which we will refer to as off-dispersion EAWs.   The purpose of this paper is to shed some light on the nature of these off-dispersion EAWs, by examining the sensitivity of the thumb curve to small deviations from the Maxwellian used in Ref.~\cite{holloway91}.

However, before starting our  analysis, we note that in a separate research thread, motivated by experiments on nonlinear
laser plasma interactions \cite{mont02},   Afeyan and collaborators \cite{afeyan04,johnston09} carried out extensive Vlasov
simulations and observed states that they referred to as  Kinetic Electrostatic Electron Nonlinear (KEEN) waves.  These large amplitude structures also exist off-dispersion; however, the  KEEN waves observed by these authors are  large amplitude, while  we focus on  relatively low amplitude EAWs, where any effect of  nonlinearity is limited to a narrow velocity range of trapped particles.
\begin{figure}
\epsfxsize=7cm \centerline{\epsffile{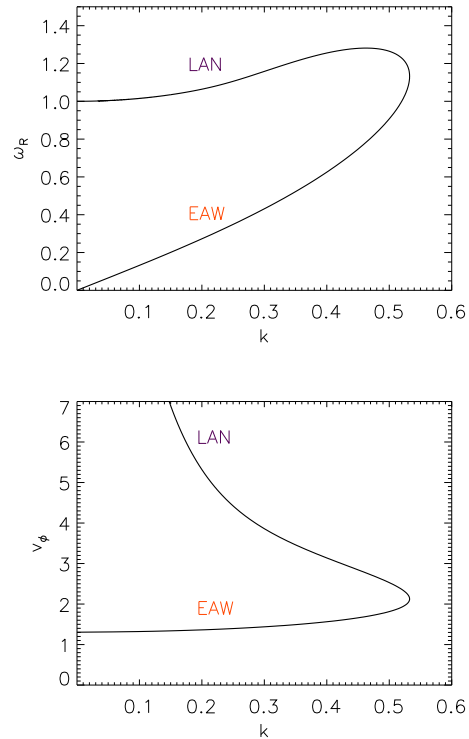}}     
\caption{Thumb curve in the  $k-\omega_{_R}$ plane displaying  branches of undamped LAN waves  and EAWs (top).
The same thumb curve plotted in the $k-v_{\phi}$ plane (bottom).} \label{fig1}
\end{figure}

Specifically, we investigate both on-dispersion and off-dispersion EAWs with a simple linear dispersion
analysis.  As noted above, there is no {\it a priori} reason the distribution in the vicinity of $v_{\phi}$  should be  Maxwellian,   since the structure  near $v_{\phi}$ is determined by the history of formation of the waves,  and  since experimentally fine details of the distribution function are difficult to measure,  it is natural to examine the alteration of the thumb curve of Ref.~\cite{holloway91} caused by  small deviations from the Maxwellian.   Of main importance here is the alteration caused by replacing the trapped particle region of the velocity distribution by  a plateau, but we also consider  the alteration caused by altering the tail of the distribution function.

For the plateau distribution function (see Eq.~(\ref{fp})) the Landau velocity integral is carefully evaluated using a high resolution trapezoidal scheme. We look for roots
of the dispersion function in the high frequency range of electron modes, treating the ions as a stationary
neutralizing background charge. Consistent with the experiments \cite{anderegg091,anderegg092} the analysis
shows that undamped EAWs exist in a wide range of the $(k,\omega_{_R})$ plane, that is, off the thumb
 curve of  \cite{holloway91}.

{}From the perspective of the linear dispersion analysis, the existence of the off-dispersion modes is easy to
understand. The Landau velocity integral in the dielectric function obtains contributions from velocities
that are well away from the plateau (the non-resonant particle contributions) and contributions from near the
plateau (the resonant particle contributions). The thumb dispersion curve is determined exclusively by
contributions from the non-resonant particles, that is, for $(k,\omega_{_R})$ on the dispersion curve, the
non-resonant contribution alone yields a dielectric function that is zero. Off the dispersion curve, the
non-resonant contribution yields a  dielectric function that is not zero, so the resonant contribution must
make up the difference, yielding a total dielectric function that is zero. Thus, for the off-dispersion
modes, electrons in the resonant (or plateau) region make a significant contribution to the mode charge
density. In that sense, the off-dispersion modes are like beam modes, for which a significant part of the
charge density resides on the beam \cite{oneil68}. As we will see, for the case of the plateau (rather than a
beam), the resonant particle charge density is associated with the two corners of the plateau.  Thus,  we call these waves
{\it corner modes}.

As one would expect, the charge density from the corners is a very sensitive function of $v_{\phi}$
in the plateau region.  Equivalently, the dielectric function has a spiky
variation for $v_{\phi}$ in the plateau region. Thus, a small change in $v_{\phi}$  can make
the resonant particle charge density have whatever value is needed to compensate for the non-zero value of the
non-resonant dielectric, yielding a total dielectric that is zero. On-dispersion EAWs are special only in
that the charge density from the two corners is equal and opposite adding to zero.  This is the case if the
phase velocity is equidistant from the two corners, that is, at the velocity mid-point of the plateau. From
this perspective, there is little difference between the on-dispersion and off-dispersion EAWs. However,
there is a significant difference between the EAWs (or corner modes) and weakly damped Langmuir waves, where
$v_{\phi}$  is well out on the tail of the velocity distribution and there is no significant
corner contribution to the mode charge density.

Another way to obtain off-dispersion waves is to alter the tail of the distribution function,  which  may not be  known precisely.  Although the dispersion curve is not as sensitive to this kind of deviation, one can ascertain systematic shift of roots off of the thumb curve.   In particular, we can show analytically that a fattening of the tail of the distribution shifts roots toward lower $k$-values and chopping the tail shifts them toward  higher  $k$-values.  Chopping produces perturbed corner charge and this idea leads to a derivation of a rule of thumb for assessing shifts caused by  general plateau type of equilibrium distribution functions.

The paper is organized as follows. In Section II we numerically analyze the roots of the electrostatic
dielectric function and discuss the wave dispersion relation for a velocity distribution flattened in a small
velocity interval.  This is followed by an analysis of the consequences  of altering the tail and the derivation of the rule of thumb.  In Section III the numerical results of Vlasov-Poisson simulations are presented and compared to the analytical predictions of Section II for the plateau distributions. Summary and Conclusions are given in Section IV.
\begin{figure}
\epsfxsize=7.cm \centerline{\epsffile{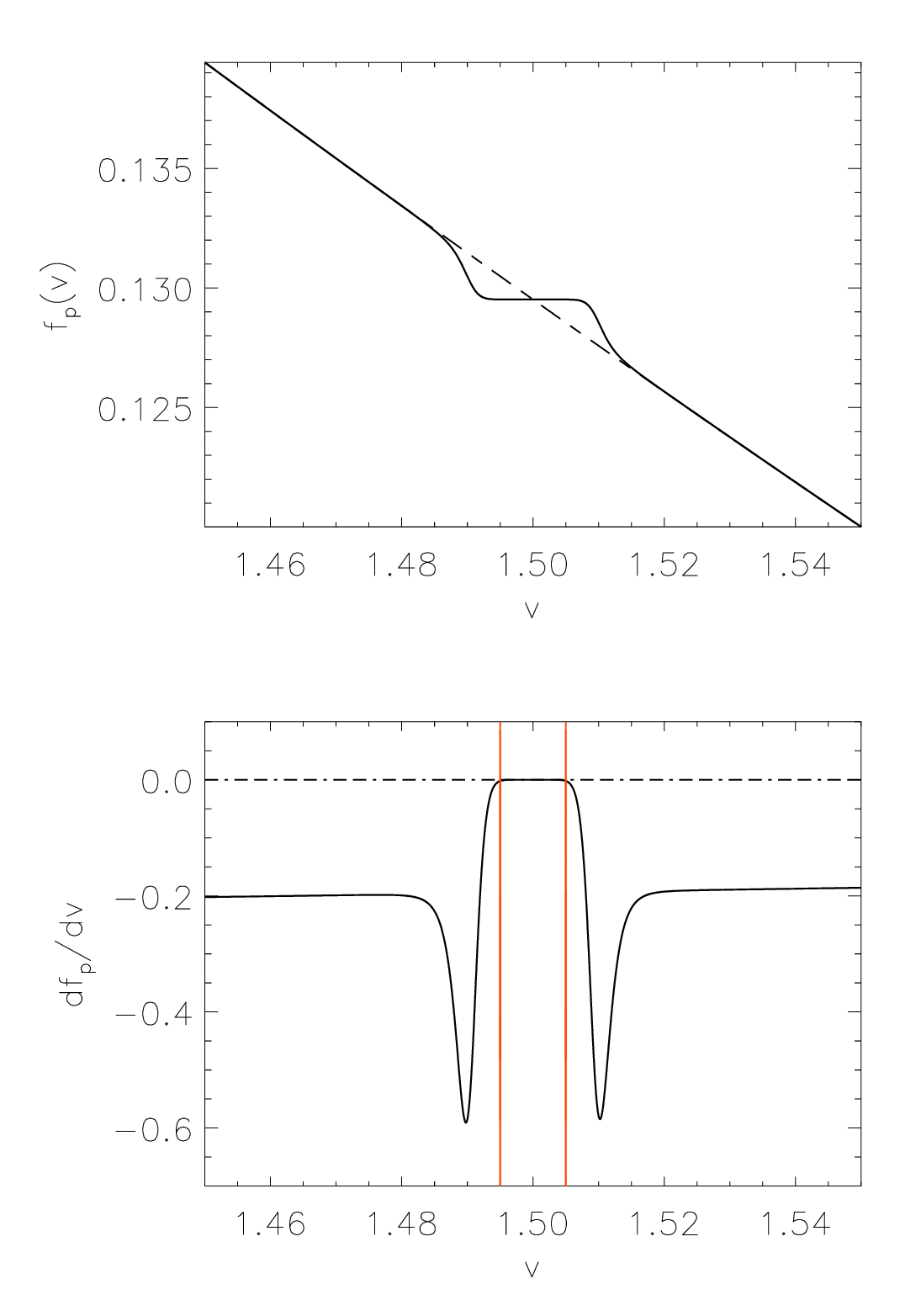}}        
\caption{Velocity dependence of the function $f_p$ (top) and  its first velocity derivative $df_p/dv$
(bottom) in the region near $V_0=1.50$. Dashed line of  top plot is the usual
Maxwellian, $f_{_M}$; red-vertical lines of the bottom plot mark the edges of the interval $[V_0-\Delta
V_p/2,V_0+\Delta V_p/2]$.} \label{fig2}
\end{figure}

\section{Wave dispersion relation}
\label{dispersion}

The propagation of electrostatic waves in a collisionless unmagnetized plasmas can be described by a  simplified 1+1+1 (one space,  one velocity, and one time dimension) Vlasov-Poisson system, which is given in dimensionless form as follows:

\begin{equation}
\label{eqVP}
\frac{\partial f}{\partial t}+ v\frac{\partial f}{\partial
x}-E\frac{\partial f}{\partial v}  = 0\,, \qquad  \frac{\partial
E}{\partial x}= 1-\int f dv\,,
\end{equation}
where $f=f(x,v,t)$ is the electron distribution function and $E=E(x,t)$ the electric field. In (\ref{eqVP}), the ions are a neutralizing background of constant density $n_0=1$, time is scaled by the inverse electron plasma frequency $\omega_p^{-1}$, velocities by the electron thermal speed $v_{th}$,  and lengths by the electron Debye length $\lambda_D$.  For   simplicity,  all the physical
quantities will be expressed in these characteristic units.

By linearizing Eqs.~(\ref{eqVP}) and following the Landau prescription \cite{landau46,krall86} for weak
wave damping, the time asymptotic solution for the complex frequency of the fluctuations
($\omega=\omega_{_R}+i\omega_{_I}$) can be obtained by looking for the roots of the dielectric
function $D(k,\omega)\simeq D_{_R}(k,\omega_{_R})+iD_{_I}(k, \omega_{_R})+i\omega_{_I}\partial
D_{_R}(k,\omega_{_R})/\partial\omega_{_R}$, where
\begin{equation}
\label{realdie}
D_{_R}=1-\frac{1}{k^2}
\dashint\! dv\, \frac{f_0'}{v-v_\phi}\,, \qquad  D_{_I}=-\frac{\pi}{k^2}  \left.f_0'\right|_{v_\phi}\,.
\end{equation}
Here, $f_0'(v) :=\partial f_0/\partial v$,   $\dashint$ indicates an integral over all $v\in\R$ with  the singularity  handled by taking the  Cauchy principal value,   $f_0$ is  the  equilibrium velocity distribution of electrons,
and $v_\phi=\omega_{_R}/k$ is the wave phase speed. The roots of $D_{_R}$ give   the real part of the wave frequency $\omega_{_R}$, while the imaginary part   is given by $\omega_{_I}=-D_{_I}/(\partial D_{_R}/\partial\omega_{_R})$.

Undamped waves can be obtained from  Eqs.~(\ref{realdie}) by  assuming $f_0$  has a velocity plateau of vanishing velocity width  at $v=v_\phi$. This renders $D_{_I}(v_{\phi})=0$ and  solution of $D_{_R}=0$ yields
$\omega_{_R}=\omega_{_R}(k)$.  In  Ref.~\cite{holloway91},  the equation $D_{_R}=0$ was solved by assuming Maxwellian  $f_0$ with the  velocity plateau of vanishing width, leading to the so-called {\it thumb curve},  the
dispersion diagram displayed in the top plot of Fig.~\ref{fig1}.   The upper branch of  this
$k-\omega_{_R}$ diagram represents Langmuir (LAN) waves, while modes of the  lower branch are usually referred to
as EAWs, since for small wavenumbers on the lower branch  $\omega_{_R}^{^{(EAW)}}\approx 1.31 k$,  which is reminiscent of acoustic  waves. In the bottom plot of Fig.~\ref{fig1},  the same thumb curve is displayed in the $k-v_{\phi}$ plane.
\begin{figure}
\epsfxsize=7cm \centerline{\epsffile{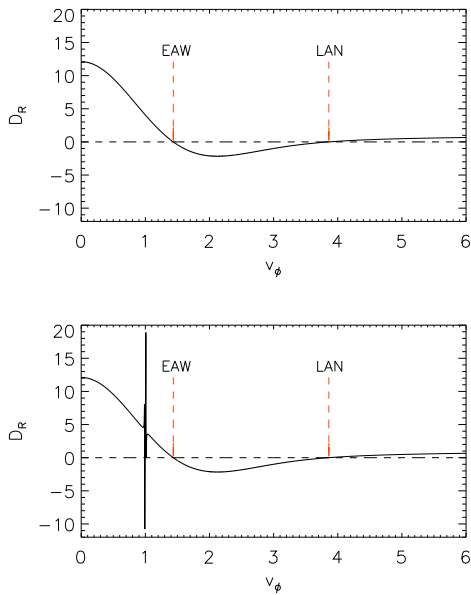}}   
\caption{Real part of the dielectric function $D_{_R}$ as a function of $v_\phi$ for $k=0.3$ for  a
Maxwellian equilibrium velocity distribution (top) and for the equilibrium distribution $f_p$
 given in Eq.~(\ref{fp}) for $V_0=1$ (bottom).} \label{fig3}
\end{figure}

From the plots in Fig.~\ref{fig1} it is evident that  no undamped roots of $D_{_R}$ exist beyond a critical
value of the wavenumber $k^*\simeq 0.53$. The presence of this nose-like structure at
$k\simeq k^*$ appears unphysical, since the group velocity of the wave at $k^*$ seems to diverge. To
understand this apparent paradox one should bear in mind that the thumb curve  does not
represent a usual dispersion relation: each point in the $(k,\omega_{_R})$ plane along the
thumb curve corresponds to a different particle velocity distribution function.   In order to get
undamped solutions, the  location of the infinitesimal  plateau in the electron velocity distribution must
slide along and always fall at $v=v_\phi$. This corresponds to changing the shape of the velocity
distribution  at  each $v_\phi$.

As  discussed in Sec.~\ref{intro}, the existence of the EAW branch has been reproduced in numerical
simulations and observed in  experiments on nonneutral plasmas.  In the experiments,  stable oscillations also  are observed off-dispersion, i.e., off  of the usual  thumb  curve.   In order to obtain insight  for understanding this experimental behavior,  we  will analyze the roots of $D_{_R}$ for an  equilibrium particle velocity distribution function that deviates from Maxwellian  by a small,  but not infinitesimal,  velocity plateau of width $\Delta V_p$  located at $v=V_0$.  Specifically, we chose the plateau distribution function  given by
\begin{equation}\label{fp}
 f_p(v)=N\left(f_{_M}(v)-\frac{f_{_M}(v)-f_{_M}(V_0)}{1+[(v-V_0)/\Delta V_p]^{n_{p}}}\right)\,,
\end{equation}
where $f_{_M}=\exp{(-v^2/2)}/\sqrt{2\pi}$ is the usual Maxwellian, $n_p$ is an even integer
(here  $n_p=10$),  $\Delta V_p=0.01$, and $N$ is a normalization constant that deviates slightly from unity.   It is worth noting that $f_p$ is  smooth in   $v$  with   derivatives  up to order $n_p$ that vanish at  $v=V_0$.

The   distribution $f_p$ and its first  derivative $df_p/dv$  are shown in Fig.~\ref{fig2},  in the  region near  $V_0$ (where  $V_0=1.5$ for illustrative purposes); the dashed line in the top plot represents the function $f_{_M}$. The red-vertical lines in the bottom plot indicate the width of the plateau $\Delta V_p$. It is clear from this figure  that in the interval
$[V_0-\Delta V_p/2,V_0+\Delta V_p/2]$ the first velocity derivative of $f_p$ obtains very small values.

We investigate the possibility of getting undamped (or weakly damped) plasma oscillations with $v_\phi$ in
the interval $[V_0-\Delta V_p/2,V_0+\Delta V_p/2]$. In this velocity interval, one can numerically calculate
the value of $D_{_R}$ for a fixed $k$, with  the imaginary part of the dielectric function being negligible because of the plateau of width $\Delta V_p$ at $v=V_0$ (see the bottom plot of  Fig.~\ref{fig2}).
\begin{figure}
\epsfxsize=7cm \centerline{\epsffile{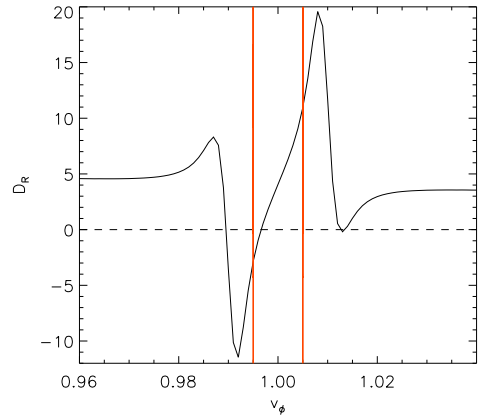}}   
\caption{A zoom of the bottom plot of Fig.~\ref{fig3} near  $v_\phi=V_0=1$.} \label{fig3bis}
\end{figure}

To calculate the value of $D_{_R}$, we compute the Cauchy principal value of the  integral of $D_{_R}$  of Eqs.~(\ref{realdie}) on a uniform velocity grid with  a standard trapezoidal scheme. The limits of  numerical  integration are set to  $|v_{_{max}}|=6$, with the distribution set  to zero outside this interval. To smoothly resolve the small plateau of width
$\Delta V_{p}=0.01$ in $f_p$,  in this interval we use  a large number of grid points
$N_v=12000$,  so as to have $\Delta v=2|v_{_{max}}|/N_v=0.001 < \Delta V_{p}$.

The two plots of Fig.~\ref{fig3} depict  the dependence of  $D_{_R}$ on $v_\phi$,  for a fixed value of the wavenumber $k=0.3$.  In the top plot we show the case where  $f_0$ is  Maxwellian, results which are essentially equivalent
to the thumb curve of Fig.~\ref{fig1}.  In the bottom plot of Fig.~\ref{fig3} we show the results obtained when the
equilibrium  distribution is chosen to be $f_p$ of Eq.~(\ref{fp}) with a small plateau at $v=V_0=1$.  The curve in the top plot displays two roots, the EAW   with $v_\phi^{^{(EAW)}}\simeq 1.44$ and the LAN wave  with $v_\phi^{^{(LAN)}}\simeq 3.86$, both being  undamped since $D_{_I}$ is assumed to vanish at each $v_{\phi}$.  Note, the curve in the bottom
plot reveals new features:  the  two roots  corresponding to the  EAW and LAN wave (in agreement with those of  the top plot)  now undergo Landau damping (very strong for the EAW), since the location of the  velocity plateau in $f_p$ is now fixed at $v=V_0=1$.  In addition, $D_{_R}$ displays a marked spike in the region around $v_\phi\simeq V_0=1$.

In Fig.~\ref{fig3bis} we zoom in for a close-up of the spike  region around $V_0$ and  find  several extra roots of $D_{_R}$. Among these roots, we focus on the one that falls within the interval $[V_0-\Delta
V_p/2,V_0+\Delta V_p/2]$ marked by the red-vertical lines, since the other roots outside this  interval are very strongly Landau damped.   Now the search for the undamped  root of $D_{_R}$ and   related analysis can be
restricted to a limited and very small region of the velocity domain, and we can increase the velocity resolution
by a factor of $100$ so as  to determine more precisely the location of the root of $D_{_R}$ in the  interval $[V_0-\Delta V_p/2,V_0+\Delta V_p/2]$. By doing this, we find that the root of $D_{_R}$ is located at $v_\phi=v_\phi^*\simeq 0.9966$.  We also evaluated  the corresponding imaginary part of the dielectric function from the second of Eqs. (\ref{realdie}), getting a small value $D_{_I}\simeq 2\times 10^{-3}$.  Moreover,  using  $\omega_{_R}=kv_\phi^*$ and  $\omega_{_I}=-kD_{_I}/(\partial D_{_R}/\partial v_\phi)|_{v_\phi^*}$,  the ratio    $R_{th}:=|\omega_{_I}/\omega_{_R}|\simeq 10^{-6}$, meaning
that this is an almost undamped solution. It is interesting to point out that if one chooses the value of
$V_0$ (the location of the plateau in $f_p$)  so that for a fixed $k$ it falls exactly on
the thumb curve of Fig.~\ref{fig1} (i.e.,  it falls exactly on the LAN or EAW branch), the roots of
$D_{_R}$ are found exactly at $v_\phi=V_0$ and they are completely undamped, since
$(df_p/dv)_{v=v_\phi=V_0}=0\Rightarrow D_{_I}=0$.

In order to establish the domain of parameters for  which electrostatic waves can exist without being  Landau
damped,  by including small plateaus in the equilibrium velocity distribution, we calculate the minimum value of
$|D_{_R}|$ in the velocity interval $[V_0-\Delta V_p/2,V_0+\Delta V_p/2]$ for different values of $k$ and
$V_0$.  Then $\min{\{|D_{_R}|\}}=0$ corresponds to a root of $D_{_R}$.

\begin{figure}
\epsfxsize=8cm \centerline{\epsffile{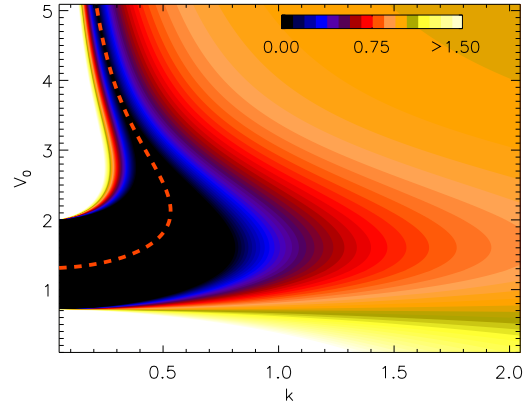}}   
\caption{Contour plot of $\min{\{|D_{_R}|\}}$ in the $k-V_0$ plane; red-dashed line is the thumb
curve.} \label{fig4}
\end{figure}

The results for $\min{\{|D_{_R}|\}}$ are displayed in the $k-V_0$ contour plot of Fig.~\ref{fig4}. Here, the
dark area represents the region where $\min{\{|D_{_R}|\}}=0$, which  corresponds to weakly damped solutions, while  outside this region $\min{\{|D_{_R}|\}}>0$, which means no solutions exist.  The red-dashed line represents the thumb curve previously shown
in the bottom plot of Fig.~\ref{fig1}.  For the dark region of the contour plot in Fig.~\ref{fig4}, for which
the roots of $D_{_R}$  are in the interval $[V_0-\Delta V_p/2,V_0+\Delta V_p/2]$, one can
evaluate the  ratio  $R_{th}$  to ascertain the importance of Landau
damping for each solution. The maximum value of this ratio in the dark region of the contour plot in Fig.~\ref{fig4} is $R_{th}^{^{max}}:=\max{\{|\omega_{_I}/\omega_{_R}|\}}\simeq 6\times 10^{-5}$, with
$|\omega_{_I}/\omega_{_R}|$ being exactly null on the  thumb curve (red-dashed line in Fig.~\ref{fig4}).
Therefore, Fig.~\ref{fig4} shows that almost undamped oscillations can be obtained with a  small
plateau in the equilibrium velocity distribution in an unexpectedly wide region around the thumb curve.  More
importantly, undamped solutions can be found well beyond the critical wavenumber $k^*$ predicted by the thumb
curve, suggesting an avenue for  understanding  the experimental results with nonneutral plasmas discussed in
Refs. \cite{anderegg091,anderegg092}.

\begin{figure}
\epsfxsize=7.cm \centerline{\epsffile{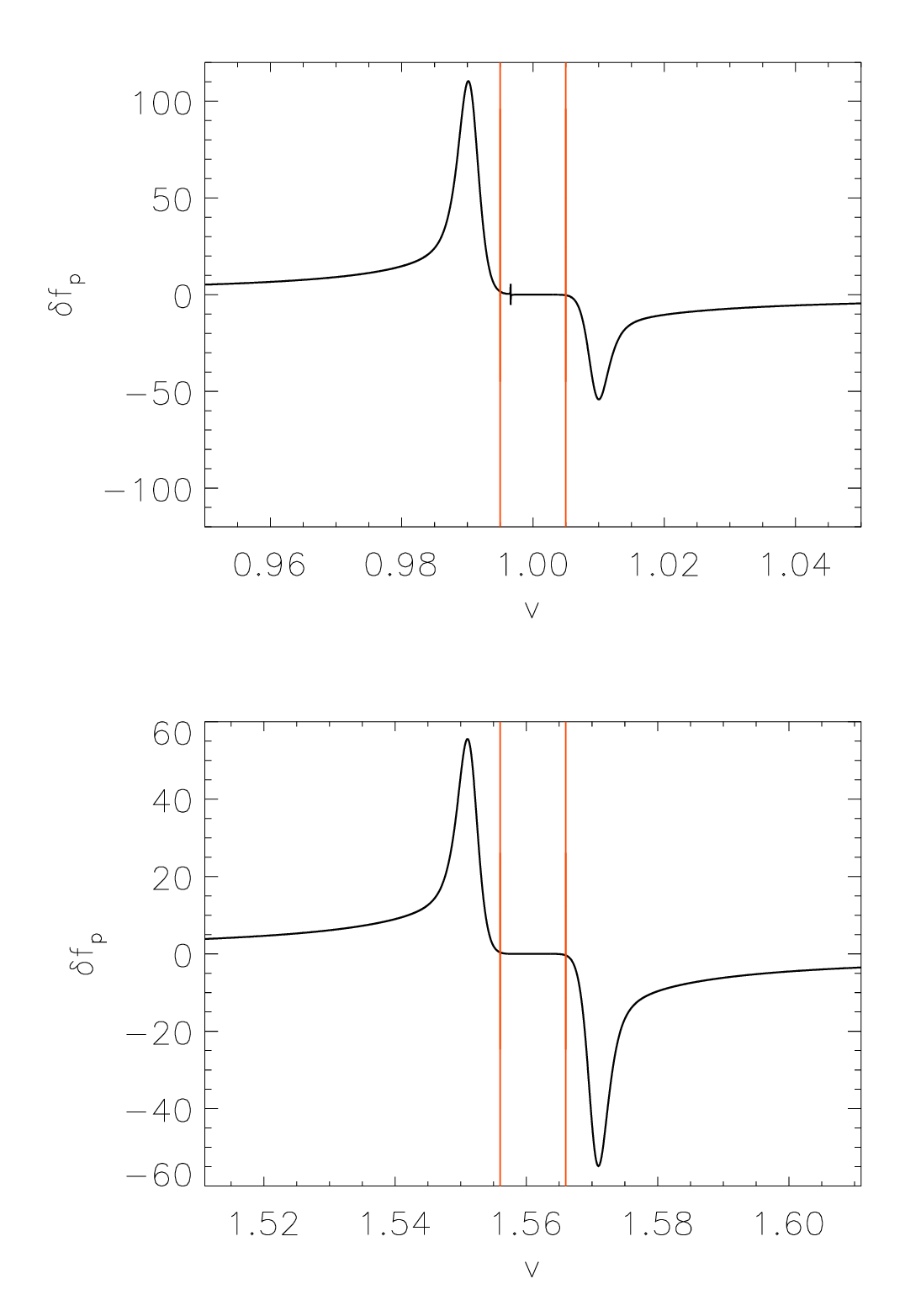}}      
\caption{Velocity dependence of the perturbed distribution $\delta f_p$ near  $v=V_0$,
for a mode with $k=0.3$ and $V_0=1$ that lies off of the thumb curve (top);   for a mode with $k=0.4$
and $V_0=1.561$ lying on the thumb curve (bottom).}\label{fig4bis}
\end{figure}

To complete our analysis, we analyze the perturbed distribution function of these undamped
oscillations, the form of which is  given by  $\delta f_p=f'_p/(v-v_\phi^*)$ \cite{krall86,shadwick94}.  We first assume
$k=0.3$ and  $V_0=1$ (as in Fig.~\ref{fig3}) and $v_\phi^*=0.9966$, which corresponds to a mode that is located off the
thumb curve (see Fig.~\ref{fig1}).   In the top plot of Fig.~\ref{fig4bis},   $\delta f_p$ is plotted as a
function of $v$ in the region around $V_0=1$. The red-vertical lines mark the interval $[V_0-\Delta
V_p/2,V_0+\Delta V_p/2]$ and a small spike is seen at the location of the  pole at $v=v_\phi^*=0.9966$.
In addition,  two pronounced peaks, visible within a velocity interval $I_v$ of width $\sim 0.04$ around $V_0$,
correspond to the sharp corners at the boundaries of the plateau.  For this off-dispersion mode,  the contributions to $\delta f_p$ due to  the two corners are not symmetric, because the wave phase speed is not in the center of the velocity plateau, $V_0$. This
means that for these off-dispersion modes,  when   $\delta f_p$ is integrated over  $I_v$ there will be a net contribution from the corners to the charge density.

The situation is different for modes that fall on the thumb curve.  The bottom plot of Fig.~\ref{fig4bis}
displays $\delta f_p$ for an on-dispersion mode with $k=0.4$ and $V_0=1.561$. As discussed
previously, when the values of $k$ and $V_0$ are such that the mode  falls  on the thumb curve, then $v_\phi$ exactly equals $V_0$, the center of  the velocity plateau.  This suppresses the  pole in the perturbed distribution (which  is valid for any   distribution with a plateau at $v=V_0$, whose first and second velocity derivatives vanish at $v=V_0$). Indeed, the Landau pole is
not visible in the bottom plot of Fig.~\ref{fig4bis} and now the contributions from the two corners  are  exactly symmetric (but of opposite sign).  Consequently, the peaks corresponding to the two corners cancel upon integration  over  $I_v$, yielding a negligible contribution to the charge density.

In reality, the tail of $f_p$ of Eq.~(\ref{fp}) is composed of a part that is  algebraic and a part that is Maxwellian.   This  deviation from Maxwellian only matters for large velocities, and it is  insignificant for  our simulations of Sec.~\ref{simulations}, where the tail is actually chopped at large velocities.  However, it does raise the question of how altering the tail affects the thumb curve.    To address this kind of deviation  consider an equilibrium  of the form $f_{0}(v)= N_1 f_{_M} + N_2 f_{\rm tail}$, where $N_1$ and $N_2$ are  yet to be determined normalization constants, and assume
\begin{equation}
f_{\rm tail} = \left\{ \begin{array}{cc}
			f_{_M} & \quad {\rm for}\ \ v\leq v_* \\
			\phi_{\rm tail}& \quad \ {\rm for}\ \ v> v_* \,,
		 \end{array} \right.
 \end{equation}
 where  $v_*>v_{\phi}>0$, and we assume  continuity at $v_*$, but smoothness is not essential.    Normalization requires
  \bq
  1= N_1 + N_2 +N_2\,  \Delta(v_*) :=N_1 + N_2 +N_2 \int^{\infty}_{v_*}\!dv \, \left(\phi_{\rm tail}-f_{_M}\right)
  \,.
  \label{D}
  \eq
 Then,  the thumb curve is given by
 \bq
  k^2=\dashint \!dv \, \frac{f_{_M}'}{v-v_{\phi}}=:\wi(v_{\phi})\,,
  \eq
  as plotted in the bottom panel of Fig.~\ref{fig1}, and its deviated form with $f_{0}(v)$  is given by
  \bq
  k^2=(N_1 + N_2) \wi + N_2 \, T(v_*)\,
  \label{T}
  \eq
 where
 \bq
 T(v_*,v_{\phi}):= \int_{v_*}^{\infty}\!\! dv\,  \frac{\phi_{\rm tail}'-f_{_M}'}{v-v_{\phi}}\,.
 \eq
 Equations  (\ref{D}) and (\ref{T}) define a one parameter family of deviated thumb curves given by
  \bq
  k^2= \wi(v_{\phi})+ N_2\big(T(v_{\phi},v_*)- \wi(v_{\phi}) \,  \Delta(v_*)\big)\,,
  \label{newX}
  \eq
 where $N_2$  determines the fraction of particles in the non-Maxwellian part of the tail of the distribution.

 We  consider  two cases: fat tails and chopped tails.  For both we suppose  $v_*>>v_{\phi}$, so to good approximation
   \bq
 T\approx  \int_{v_*}^{\infty}\!\! \frac{dv}{v}\,  \left(\phi_{\rm tail}'-f_{_M}'\right) \,.
 \label{Tapprox}
 \eq
 \noindent \underline{Fat Tails}:   A fat tail is one where $\phi_{\rm tail}'>f_{_M}'$, which is the case if  $\phi_{\rm tail}$ is  a kappa-distribution function which for large $v$ behaves as  $\phi_{\rm tail}\sim c/v^{\alpha}$,   where $c>0$ is a constant.  Then
 \bq
 T\approx-\frac{\alpha}{\alpha +1}\,  \frac{c}{v_*^{\alpha + 1}}\,,\qquad
  \Delta\approx \int_{v_*}^{\infty}\!\! dv\,  \phi_{\rm tail} \approx \frac{1}{\alpha - 1} \, \frac{c}{v_*^{\alpha - 1}}\,,
\eq
and  (\ref{newX}) becomes
  \bq
  k^2= \wi+ \frac{c N_2}{v_*^{\alpha + 1}} \left(- \frac{\alpha}{\alpha + 1} + \wi \, \frac{v_*^2}{ 1-\alpha}\right)
  \label{FatnewX}
  \eq
 {\it Therefore, for $\alpha >1$, which physically is clearly desired, the new contribution is negative and the thumb curve moves so as to decrease $k^2$.}  This means fattening the tail with $v_*>>v_{\phi}$ shifts the thumb curve upwards in the bottom plot of Fig.~\ref{fig1}.
 \medskip

\noindent\underline{Chopped Tails}:  For thin tails the situation is not so clear cut.   First, to remove the Maxwellian tail we set $N_1=0$, which amounts to setting $f_0=N_2\big(H(v_*-v)f_{_M} + H(v-v_*)\phi_{\rm tail}\big)$,  where $H$ is the Heaviside function that is unity for positive argument and zero for negative.   For thin tails,   $\Delta$ would be negative, which would tend to move $k^2$ toward larger values, but the $T$ term is more subtle.    As an extreme case we will chop the tail at $v_*$ by setting $\phi_{\rm tail}\equiv0$, giving the derivative of $f_0$ a jump discontinuity.  This jump can be removed, e.g.\  by   interpolation with a steep slope, but the results do not change much.   With the chopped choice
  \bq
  \De(v_*)=- \int_{v_*}^{\infty}\!\!dv\, f_{_M}\,,
  \label{Dchop}
  \eq
 making the last term of (\ref{newX}) positive, as opposed to the case of the fat tail. For $T$ we must evaluate the jump using
  $f_{0}'= N_2\big(-\delta(v-v_*)f_{_M} +H(v_*-v) f_{_M}'\big)$.  With some manipulation, we obtain
  \bq
   k^2= \wi+ \frac{N_2}{\sqrt{2\pi}}\left( - \frac{e^{-\frac{v_*^2}{2}}}{v_*}
    + (1+\wi) \int_{v_*}^{\infty}\!\!dv\,  e^{-\frac{v^2}{2}}\right)\,,
   \eq
  and it remains to determine which of the `correction terms' within the parentheses dominates.
  To this end we make use of the following inequality \cite{AS}:
  \bq
 \frac{2\,  e^{-\frac{v_*^2}{2}}}{v_*+\sqrt{v_*^2 +4}}    < \int_{v_*}^{\infty}\!\!dv\,  e^{-\frac{v^2}{2}}\,,
 \qquad v_* >0\,.
 \eq
  Evidently, $k^2$ increases if
  \bq
  -\frac1{v_*} +  \frac{2(1+\wi)}{v_*+\sqrt{v_*^2 +4}}   >0
  \eq
  and a simple calculation shows this is true if
  \bq
  v_*^2>\frac{1}{\wi(1+\wi)}\,.
  \eq
 Examination of  Fig.~3 of Ref.~\cite{holloway91}  reveals that $\wi\gtrsim 1$ which means for large $v_*$,  {\it a chopped tail  shifts the thumb curve so as to increase the  values of $k^2$.}   This means chopping  the tail at $v_*>>v_{\phi}$ shifts the thumb curve downwards in the bottom plot of Fig.~\ref{fig1}.

 One can interpret the chopping of the tail as contributing an extreme kind of corner charge to the perturbed charge distribution.  In closing this section we will use this idea to obtain a rule of thumb for explaining frequency shifts due to plateaus.  Consider  the following plateau distribution with extreme corners:
\bqy
f_{ep}&=& N_{_M} f_{_M} \left[H(v - v_+)+ H(v_- -v) \right)]\nonumber\\
&{\ }&\hspace{.1 cm}
+ N_p \left[H(v- v_-)- H(v-v_+) \right)]\,,
\label{fe}
\eqy
i.e., with jump discontinuities at $v_{\pm}:=V_0\pm\Delta V/2$.
Normalization of (\ref{fe}) requires
\bq
1=  N_{_M}  -N_{_M}  \int_{v_-}^{v_+}\!\!dv\,   f_{_M}
+ N_p\,  \Delta V\,.
\label{norm2}
\eq
Differentiating of (\ref{fe})  gives
\bqy
f_{ep}'&=& N_{_M} f_{_M}' \left[H(v-v_+)+ H(v_--v) \right)]
\nonumber\\
&{\ }&\hspace{.1  cm}
(N_{_M} f_{_M}-N_p) \left[
 \delta(v-v_+) - \delta(v-v_-)
\right]
\label{fepp}
\eqy
where the delta function terms represent the corner contributions.  Upon inserting (\ref{fepp})  into (\ref{realdie}),  setting $D_{_R}=0$,  making use of (\ref{norm2}), and manipulating, we obtain
\bqy
k^2
&=&
 M +M  \left(N_{_M}  \int_{v_-}^{v_+}\!\!dv\,   f_{_M}
-  N_p\,  \Delta V\right)
  \nonumber\\
&{\ }& \hspace{.1 cm}  -   N_{_M}\int_{v_-}^{v_+} \!dv\,   \frac{f_{_M}'}{v-v_{\phi}}
  \nonumber\\
&{\ }& \hspace{.1 cm}
+\frac{ N_{_M} f_{_M}^{(+)}-N_p}{v_+-v_{\phi}}  - \frac{ N_{_M} f_{_M}^{(-)}-N_p}{v_--v_{\phi}}\,,
\label{dispp}
\eqy
where $f_{_M}^{(\pm)}:=f_{_M}(v_{\pm})$.
Expression (\ref{dispp}) is valid if $f_{_M}$ is replaced by any homogeneous equilibrium distribution function.   Now expanding  in  $\Delta V/V_0<<1$, retaining the leading order,  and assuming  $v_- < v_{\phi}<v_+$,   to avoid Landau damping,  produces
\bq
k^2 \approx  M(v_{\phi}) +\frac{f_{_M}^{(+)}-N_p}{v_+-v_{\phi}}  - \frac{f_{_M}^{(-)}-N_p}{v_--v_{\phi}}\,,
\label{dispApprox}
\eq
an expression that displays the  two corner charge corrections, which have opposite signs provided $f_{_M}^{(-)}>N_p> f_{_M}^{(+)}$ and $v_-< v_{\phi}< v_+$.   The direction of the shift in $k^2$ depends on which dominates.  From Eq.~(\ref{dispApprox}) we obtain the following compact {\it rule of thumb}:
\bq
k^2=M + \frac{(V_0-v_{\phi})\left(f_{_M}^{(+)}-f_{_M}^{(-)}\right)}{(V_0-v_{\phi})^2 - (\Delta V/2)^2}\,,
\label{rot}
\eq
 which makes it very clear how the sign is determined.  Note that the corners produce a waterbag-like  denominator, as opposed to beam modes \cite{oneil68}, and this contribution vanishes for $v_{\phi}=V_0$,  in agreement with our discussion above.  Equation (\ref{rot}) can be used in a practical sense:  even though in this derivation $f_{_M}^{(\pm)}$ represents  the values of the Maxwellian  just below and just above the plateau, their difference  can be viewed as a measure of the total corner charge contributions, while $\Delta V$ serves as  an effective plateau width.  Thus, the rule of thumb provides a general rule for parameter dependencies of frequency shifts, one that should be useful for analyzing experimental data.

\begin{figure}
\epsfxsize=7cm \centerline{\epsffile{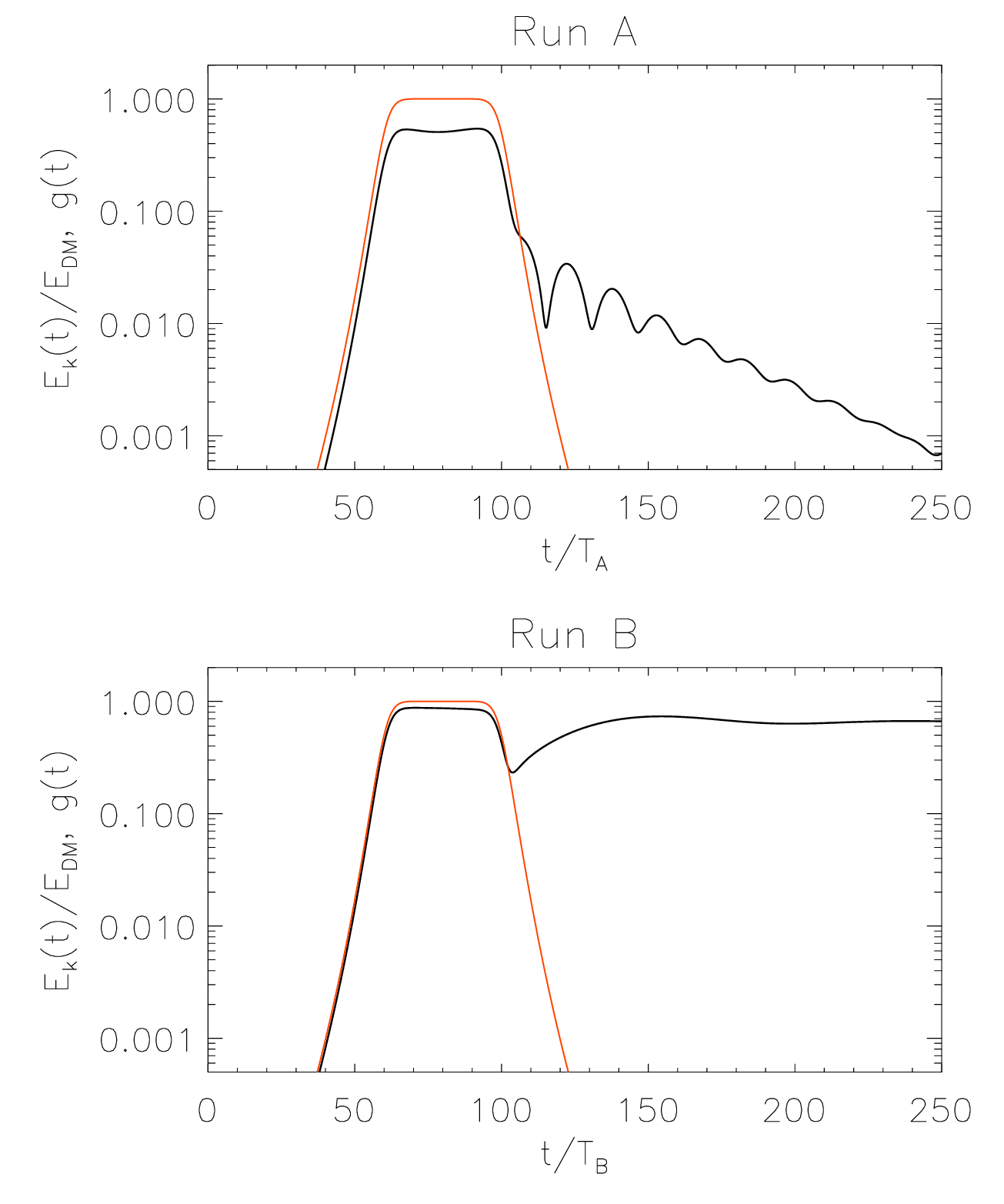}}     
\caption{Time evolution of the electric field spectral component $E_k(t)$ (normalized by the maximum driver
amplitude $E_{_{DM}}$) for Run A (top) and Run B (bottom); in both plots the red (gray) curve
represents the function $g(t)$.}\label{fig5}
\end{figure}
\section{Numerical simulations}
\label{simulations}

Because the results of Sec.~\ref{dispersion} are essentially linear in nature, we  investigate their resilience  by resorting to
simulations of the nonlinear Vlasov-Poisson system of  Eqs.~(\ref{eqVP}).  In particular, we concentrate on modes that arise from deviations of the thumb curve, off-dispersion modes,  caused by the small plateaus.

Our simulations are performed with an  Eulerian Vlasov code based on the well-know splitting time advance method given in  Ref.~\cite{knorr76}. The phase space domain for the simulations is   $\mathcal{D}=[0,L]\times[-v_{_{max}},v_{_{max}}]$.  Periodic boundary conditions in $x$ are assumed,    while the electron velocity distribution is set equal to zero for
$|v|>v_{_{max}}=6$.  We investigate disturbances near the  initial equilibrium of  Eq.~(\ref{fp}), with $n_p=10$ and $\Delta V_p=0.01$ by applying a drive force. The $x$-direction is discretized with $N_x=256$ grid points, while
$v$-direction with  $N_v=12000$.   Our goal is to numerically  analyze the modes  predicted by Fig.~\ref{fig4}.

The plasma is driven by an external electric field that is taken to be a sinusoidal traveling wave  with
phase speed $v_{\phi_{_D}}$ that exactly matches $V_0$, the location  of the plateau of $f_p$.  The
explicit form of the external field is
\begin{equation}\label{driver}
 E_{_D}(x,t)=g(t)E_{_{DM}}\sin{(kx-\omega_{_D} t)}\,,
\end{equation}
where $E_{_{DM}}$ is the maximum driver amplitude, $k=2\pi/L$ is the drive wavenumber, which is the maximum
wavelength that fits in the simulation box, $\omega_{_D}=kv_{\phi_{_D}}$ is the drive frequency,  and
$g(t)=[1+(t-\tau)^n/\Delta\tau^n]^{-1}$ is a profile that determines the ramping up and ramping down of the drive.  The external electric field is applied directly to the electrons by adding $E_{_D}$ to $E$ in the Vlasov equation.   An abrupt turn-on or turn-off of the drive field would excite LAN waves and  complicate the results. Thus, we  choose $n=10$  so  $g(t)$  amounts to a nearly adiabatic  turn-on and turn-off. The driver amplitude remains near $E_{_{DM}}$ for a time interval of order $\Delta\tau$
centered at $t=\tau$ and it is zero for $t\geq t_{\rm off}\simeq\tau+\Delta\tau/2$.   We will analyze the  plasma response  for many wave periods after the driver has been turned off.


The effect of the driver is to prepare a state (i.e.\ distribution function), which is then used as an  initial condition  for the undriven Vlasov-Poisson system of Eqs.~(\ref{eqVP}).   This type of initial condition is an example  of those called {\it dynamically accessible} in  \cite{Morrison89, Morrison90,Morrison92}, where they were advocated and discussed in detail.  Ultimately,  any perturbation of a known distribution function within the confines of Vlasov-Poisson theory must, in fact,  be caused by an electric field,  since there are no other forces available. Thus it is physically very natural to consider such initial conditions.  Dynamically accessible initial conditions are also important because they have a Hamiltonian origin and, consequently,  preserve phase space constraints.   Because the perturbed distribution function is obtained by evaluating the known state on particle orbits, the perturbed distribution function must have the same level set topology as the unperturbed and the areas between any level set contours must be preserved.  For our simulations, the dynamically accessible initial conditions used  amount to evaluating the plateau distribution of Eq.~(\ref{fp}) on the orbits (run backwards) produced by the total electric field in the interval $0<t<t_{\rm off}$.  Thus, the initial condition for the undriven dynamics that begins at $t_{\rm off}$ is a symplectic rearrangement of $f_p$ \cite{Morrison92,hagstrom,Morrison12}.

\begin{figure}
\epsfxsize=7cm \centerline{\epsffile{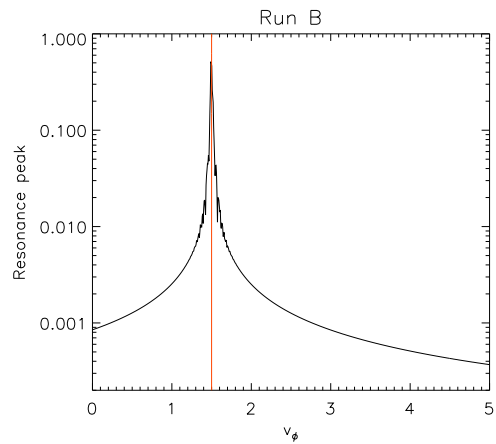}}     
\caption{Resonance peak for Run B; red-vertical line indicates the value of the driver phase
velocity $v_{\phi_{_D}}$.}\label{fig7}
\end{figure}

We performed $N_{sim}=64$ simulations for various values of  the plateau velocity position $V_0$ of  $f_p$ and the
wavenumber $k$, in order to numerically investigate the  predictions of Fig.~\ref{fig4}.
For each simulation we chose $\Delta\tau=20 T$, where $T=2\pi/kv_{\phi_{_D}}$ is the wave period (that is the
plasma is driven for 20 wave periods) and $\tau=80 T$. Typically the maximum time for the simulations is $t_{max}=250 T$, but when  needed the system evolution is followed up to  $1000 T$.  The  driver amplitude is set for
each simulation by adjusting the driver trapping time $\tau_{trap}:=2\pi/\sqrt{kE_{_{DM}}}$ to be larger than $t_{max}$. In
particular,  we set $\tau_{trap}=125\Delta\tau=10t_{max}$, so that trapping does not play a large role in the system evolution, i.e.~the distribution function changes little during the simulation.

Numerical results for $E$,  show two different kinds of electric field response for $t>t_{\rm off}$. To see this we plot $E_k(t)$, the  electric field $k$-spectral component, in the semi-log plots of Figs.~\ref{fig5}, for two different runs  denoted by A and B with
parameters given in Table \ref{tab1}.

\begin{table} [h]
 \begin{center}
\begin{tabular}{c c c c c c c c c c c c}
\hline
\bfseries Run & & & & & \bfseries $k$ & & & & &\bfseries $v_{\phi_{D}}$ \\
\hline
\hline
A & & & & & 0.9 & & & & & 0.3 \\
\hline
B & & & & & 0.7 & & & & & 1.5 \\
\hline
C & & & & & 0.4 & & & & & 1.561 \\
\hline
\end{tabular}
\caption{Relevant parameters for Runs A, B and C}
\label{tab1}
\end{center}
\end{table}

In the plots of Figs. \ref{fig5}, the electric signals are normalized by the corresponding maximum driver
amplitude $E_{_{DM}}$ and the red (gray) line represents the function $g(t)$. As is evident from the plots, in Run A we observe damped (more or less exponentially) oscillations after the driver has been turned off,  consistent with Landau damping, while in Run B we observe a stable electric response, consistent with plateau suppression. Figure \ref{fig7} is a semi-log plot of  the  spectral electrostatic energy, obtained  by the Fourier analysis of $E$ for $t> t_{off}$, as a function of $v_\phi$,   for the case of Run B, where a stable plasma response is recovered at $t>t_{off}$.  This plot reveals that the electric field propagates with a phase velocity $v_\phi$ near the driver phase velocity $v_{\phi_{D}}$ (red-vertical line).
\begin{figure}
\epsfxsize=7cm \centerline{\epsffile{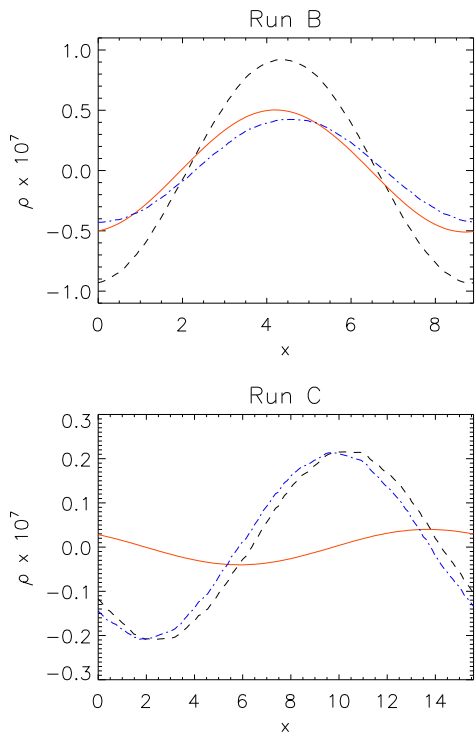}}     
\caption{Spatial dependence of the electron charge densities $\rho$ (black-dashed line), $\rho_1$ (red-solid
line),  and $\rho_2$ (blue-dot-dashed line) for Run B (top) and Run C (bottom).} \label{fig7bis}
\end{figure}

Moreover, for Run B we  evaluated the electron charge density $\rho=1-\int f dv$ at the end of the
simulation. This undamped mode   with the parameters  of Table  \ref{tab1}  is located off  of the thumb curve (see Fig.~\ref{fig1}).  The black-dashed line in the top plot of Fig \ref{fig7bis} is  the total electron charge density, $\rho$,  obtained by integrating $f$ over the  velocity interval $I=[-v_{_{max}},v_{_{max}}]$,  the red-solid line is the charge density, $\rho_1$,  obtained by
integrating $f$ over the velocity interval $I_v=[1.48,1.52]$ near  $V_0$, while the blue-dot-dashed line
is  the charge density, $\rho_2$,  obtained by  integrating over the complement of  $I_v$.  Clearly,  $\rho=\rho_1+\rho_2$.  From the  plot it is seen that  the contribution to the charge density coming from the small interval $I_v$ around $V_0$, which contains the sharp corners at the boundaries of the plateau,  is comparable to or even a bit larger than the
contribution from the rest of the  velocity distribution.  The same calculation has also been
performed for a third   Run C, whose parameters are summarized in table \ref{tab1}.   This mode falls on the thumb curve. Here, as  can be seen in the bottom plot of Fig.~\ref{fig7bis}, the contribution to the charge density from the sharp corners within the interval $I_v$ around $V_0$ (red-solid line) is significantly smaller than that of the rest of the velocity distribution. These runs provide numerical evidence that supports  the  predictions of Sec.~\ref{dispersion} for the perturbed distribution function summarized in Fig.~\ref{fig4bis}.

In order to reproduce numerically the  predictions displayed in Fig.~\ref{fig4}, we analyzed in
detail the results of $N_{sim}=64$ numerical experiments. For each simulation we evaluated the
real part of the frequency $\omega_{_R}$ and the wave damping rate $\omega_{_I}$ after the external driver was
 turned off.   Each simulation is then characterized by  calculating   $|\omega_{_I}/\omega_{_R}|:=R$ from the simulation data.  To compare the simulation results  with the contour plot of Fig.~\ref{fig4},  we use the value $R_{th}^{^{max}}$ from Sec.~\ref{dispersion}  as a threshold to divide our 64 simulations into two classes: Class $1$ for which $R\leq R_{th}^{^{max}}$
and Class $2$ for which $R>R_{th}^{^{max}}$.

\begin{figure}
\epsfxsize=8cm \centerline{\epsffile{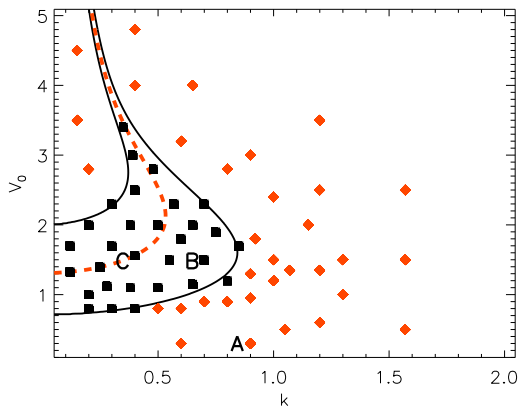}}     
\caption{$k-V_0$ scatter plot for Class 1 simulations (black squares) and Class 2 simulations (red diamonds).  The red-dashed line is  the thumb curve, while the black curves delimit  the dark
region of the contour plot of Fig.~\ref{fig4}.}\label{fig8}
\end{figure}

These results are summarized in the  $k-V_0$ scatter plot of Fig.~\ref{fig8}, where the simulations of Class
$1$ are indicated by black squares, while those of Class $2$ by red diamonds.  In this figure the red-dashed
line indicates as usual the thumb curve, while the black-solid lines  delimits  the dark region of the
contour plot in Fig.~\ref{fig4},  where almost undamped roots of the dielectric function have been recovered.
Figure \ref{fig8} clearly shows that the black squares fall within the black-solid line, while the
red diamonds lie outside this line; thus,  results of the analysis of Fig.~\ref{fig4} are well corroborated by the simulations.
The simulations of  runs Run A, Run B,  and Run C are indicated by capital letters in Fig.~\ref{fig8}.  Runs A  fall outside  the black-solid lines,  Runs B fall inside the dark region with black solid-line boundaries, and  Runs C  are exactly on the thumb curve indicated by the  red-dashed line.

\section{Summary and conclusions}

In the present work,  roots of the electrostatic dielectric function were analyzed when a velocity
plateau of small but nonvanishing width is present in the equilibrium velocity distribution of electrons. The
numerical solution of the Landau integral, performed through a high resolution scheme, allowed us to show that
quasi-undamped plasma oscillations can be obtained off of the thumb curve of Ref.~\cite{holloway91}.   By solving
numerically the Landau integral, we noted that the presence of the velocity plateau, even of very
small width, can highly affect the real part of the dielectric function, producing marked spikes within the
velocity interval around the plateau.  In a wide region of the $k-\omega_{_R}$ plane, almost undamped roots of
the dielectric function were obtained.  Examination of  the perturbed electron distribution function revealed that
most of the charge density associated with  the off-dispersion oscillations comes from the sharp
corners at the boundaries of the velocity plateau,  and for this reason we called these new modes corner modes.   A rule of thumb was derived by assuming infinitely sharp corners, a rule useful for gauging how a plateau shifts roots off of the thumb curve.

Next, these analytical predictions were  compared with the results of Eulerian Vlasov-Poisson
simulations with high resolution in velocity space.  Our simulations were initiated by applying a wave-like external electric field to drive the plateau distribution of Eq.~(\ref{fp}) off of equilibrium.   The external electric field was turned on and off adiabatically in such a way to avoid the excitation of usual Langmuir waves.   Also, the amplitude of the external driver was chosen to be very small so  trapping effects were  minimized.  As discussed in Sec.~\ref{simulations}, the numerical results of these nonlinear simulations corroborated the linear results of Sec.~\ref{dispersion}.

Although we spoke of off-dispersion results, it is important to note that all of the roots obtained in Sec.~\ref{dispersion} are actual linear electrostatic plasma oscillations: while LAN waves are approximate time asymptotic states as shown by Landau;  EAWs, corner modes, stationary inflection point modes, etc.\ are all  exact linear plasma oscillations for specific homogeneous equilibrium distribution functions.  In essence, what we are really attempting is to find a homogeneous equilibrium distribution function that best describes a weakly nonlinear theory.  Because plasmas can exist in states away from thermodynamic equilibrium for substantial lengths of time,   there is no {\it a priori} reason to believe  the distribution function is  Maxwellian, particularly if wave-like disturbances are excited.  Because of the sensitivity of the dispersion relation to equilibrium distribution functions, in the tail but particularly near $v_\phi$, we conclude that the  thumb curve of Ref.~\cite{holloway91} is of limited predictive capability.

Since the original BGK paper \cite{bernstein57}, it has been understood that there is nonuniqueness in the construction of these nonlinear modes:   a given electric field can be consistent with  a large class of distribution functions.   It is difficult to pin down the shape of the distribution function for the trapped particle population, because the formation of this distribution  depends on the time history.   This is true both in experiments and simulations,  whether the BGK modes evolve out of  instability or arise by driving the plasma as we have done here.   In the case of small amplitude disturbances, we make the point that this arbitrariness is the same as that in choosing the appropriate shape of $f_0(v)$ near $v_\phi$.  For very small disturbances, the stationary inflection point modes of \cite{shadwick94} are natural candidates. For larger disturbances, the off-dispersion modes of this paper  appear to be an attractive  alternative.

Thus, the  present work bears on  the interpretation of recent results of nonneutral plasma experiments that provide evidence for undamped off-dispersion modes.    In the experiments of \cite{anderegg091,anderegg092},  any  plateau-like structures in the electron velocity distribution are created dynamically by means of an external driver electric field that  traps resonant particles.
After the driving process, the plasma is strongly inhomogeneous, with the formation of humps and depressions in the particle distribution function, produced by the nonlinear dynamics triggered by the external field.  Thus, one would think that a linear analysis might not be relevant; however,  by designing a plateau and tail for a distribution function that has  modes that match  the experiment,  it appears that one can obtain a linear theory consistent with some of the experimental results, and possibly even infer  information about the trapped particles.  In fact, we note that the rule of thumb explains the frequency shifts observed in the    experiments of \cite{anderegg092}, but continuing with this line of investigation is beyond the scope of the present paper, so we conclude here.

\section*{Acknowledgments}
The Vlasov simulations discussed in the present paper have been run on the parallel machines at the high
performance computing center CINECA (Bologna, Italy), within the ISCRA class A project VMSP - HP10AWSJEW.  PJM was supported by U.S.~Dept.\ of Energy Contract \# DE-FG05-80ET-53088.  TMO was supported by National Science Foundation grant PHY-0903877 and Department of Energy grant DE-SC0002451.




\begin{thebibliography}{99}
\bibitem{landau46} L. D. Landau, J. Phys. (Moscow) {\bf 10},  25 (1946).
\bibitem{oneil65} T. M. O'Neil, Phys. Fluids {\bf 8}, 2255 (1965).
\bibitem{holloway91} J. P. Holloway and J. J. Dorning, Phys. Rev. A  {\bf 44}, 3856 (1991).
\bibitem{bernstein57} I. B. Bernstein, J. M. Greene and M. D. Kruskal, Phys. Rev.
{\bf 108}, 546 (1957).
\bibitem{demeio91} L. Demeio and J. P. Holloway, J. Plasma Phys. {\bf 46}, 63 (1991).
\bibitem{shadwick94} B. A. Shadwick and P. J. Morrison, Phys. Lett. A {\bf 184}, 277 (1994).
\bibitem{hagstrom} G. I. Hagstrom and P. J. Morrison, Transport Theory  Stat.  Phys. {\bf 39}, 466 (2011).
\bibitem{valentini06} F. Valentini, T. M. O'Neil and D. H. Dubin, Phys. Plasmas {\bf 13},  052303 (2006).
\bibitem{afeyan04} B. Afeyan, K. Won, V. Savchenko, T. W. Johnston, A. Ghizzon, and P Bertrand, ``Kinetic
Electrostatic Electron Nonlinear (KEEN) Waves and their Interactions Driven by the Ponderomotive Force of
Crossing Laser Beams," Proc. Inertial Fusion Sciences and Applications 2003 (B. Hamel, D. D. Meyerhofer, J.
Meyer-ter-Vehn, and H. Azechi, Eds.), Monterey: American Nuclear Society (2004) p. 213B.
\bibitem{johnston09} T. W. Johnston, Y. Tyshetskiy, A. Ghizzo and P. Bertrand, Phys. Plasmas {\bf 16},
042105 (2009).
\bibitem{valentini111} F. Valentini, F. Califano, D. Perrone, F. Pegoraro and P. Veltri,
Phys. Rev. Lett.  {\bf   106}, 165002 (2011).
\bibitem{valentini112} F. Valentini, F. Califano, D. Perrone, F. Pegoraro and P. Veltri,
Plasma Phyc. Control. Fusion  {\bf 53}, 105017 (2011).
\bibitem{gary85} S. P. Gary and R. L. Tokar, Phys. Fluids  {\bf 28}, 2439 (1985).
\bibitem{anderegg091} F. Anderegg, C. F. Driscoll, D. H. Dubin, T. M. O'Neil, Phys. Rev. Lett. {\bf 102},
09500 (2009).
\bibitem{anderegg092} F. Anderegg, C. F. Driscoll, D. H. Dubin, T. M. O'Neil and F. Valentini,
Phys. Plasmas {\bf 16},  055705 (2009).
\bibitem{mont02} S. Montgomery, J. A. Cobble, J. C. Fern�ndez, R. J. Focia, R. P. Johnson,
  N. Renard-LeGalloudec, H. A. Rose, and D. A. Russell,   Phys. Plasmas  {\bf 9}, 2311 (2002).
\bibitem{oneil68} T. M. O'Neil and J. H. Malmberg, Phys. Fluids {\bf 11}, 1754 (1968).
\bibitem{krall86} N. A. Krall and A. W. Trivelpiece, {\it Principles of Plasma Physics},
(San Francisco Press, San Francisco, CA, 1986).
\bibitem{AS}   M. Abramowitz and I. Stegun, {\it (1972), Handbook of Mathematical Functions}, (Dover Publications, New York, 1972), item  {\bf 7.1.13}  p.~298.
\bibitem{knorr76} C. Z. Cheng and G. Knorr, J. Comp. Phys. {\bf 22}, 330 (1976).
\bibitem{Morrison89} P. J. Morrison and D. Pfirsch,  Phys. Rev. A {\bf 40}, 3898 (1989).
\bibitem{Morrison90} P. J. Morrison and D. Pfirsch, Phys. Fluids B {\bf 2}, 1105 (1990).
\bibitem{Morrison92} P. J. Morrison and D. Pfirsch,   Phys. Fluids B {\bf 4}, 3038 (1992).
\bibitem{Morrison12}  P. J. Morrison,   Math-for-Industry  {\bf 39},  64  (2012).
\end{thebibliography}
\end{document}